\documentclass[aps,pra,reprint,twocolumn,superscriptaddress,showpacs,showkeys,floatfix,bibnotes]{revtex4-2}
\usepackage{polski}
\usepackage{amsmath,amssymb,amstext}
\usepackage{ulem}
\usepackage[usenames,dvipsnames]{color}
\usepackage{bm}
\usepackage{mathtools}
\usepackage[x11names]{xcolor}
\newtagform{blue}{\color{RoyalBlue3}(}{)}
\usepackage{graphicx}
\usepackage{braket}
\usepackage{natbib}
\usepackage{comment}
\usepackage{dcolumn}
\usepackage[english]{babel}
\usepackage{wasysym}
\usepackage[colorlinks,bookmarks=false,citecolor=blue,linkcolor=red,urlcolor=blue]{hyperref}

\begin{document}

\title{Transport of Vector Solitons in Spin-Dependent Nonlinear Thouless Pumps}

\author{Xuzhen Cao}
\affiliation{State Key Laboratory of Quantum Optics and Quantum Optics Devices, Institute of Laser Spectroscopy, Shanxi University, Taiyuan 030006, China}
\affiliation{Collaborative Innovation Center of Extreme Optics, Shanxi University, Taiyuan, Shanxi 030006, China}

\author{Chunyu Jia}
\affiliation{Department of Physics, Zhejiang Normal University, Jinhua 321004, China}

\author{Ying Hu}
\thanks{huying@sxu.edu.cn}
\affiliation{State Key Laboratory of Quantum Optics and Quantum Optics Devices, Institute of Laser Spectroscopy, Shanxi University, Taiyuan 030006, China}
\affiliation{Collaborative Innovation Center of Extreme Optics, Shanxi University, Taiyuan, Shanxi 030006, China}

\author{Zhaoxin Liang}
\thanks{zhxliang@zjnu.edu.cn}
\affiliation{Department of Physics, Zhejiang Normal University, Jinhua 321004, China}

\begin{abstract}
In nonlinear topological physics, Thouless pumping of nonlinear excitations is a central topic, often illustrated by scalar solitons. Vector solitons, with the additional spin degree of freedom, exhibit phenomena absent in scalar solitons due to enriched interplay between nonlinearity and topology. Here, we theoretically investigate Thouless pumping of vector solitons in a two-component Bose-Einstein condensate confined in spin-dependent optical superlattices, using both numerical solutions of the Gross-Pitaevskii equation and the Lagrangian variational approach. The spin-up and spin-down components experience superlattice potentials that are displaced by a tunable distance $d_r$, leading to a vector soliton state with a relative shift between its components. We demonstrate that $d_r$, as an independent degree of freedom, offers a novel control parameter for manipulating the nonlinear topological phase transition of vector solitons. Specifically, when $d_r=0$, both components are either pumped or arrested, depending on the interaction strength. When fixing the interaction strength and varying $d_r$, remarkably, we find that an arrested vector soliton can re-enter the pumped regime and exhibits a quantized shift. As $d_r$ continues to increase, the vector soliton transitions into a dynamically arrested state; however, with further increases in $d_r$, the quantized shift revives. Our work paves new routes for engineering nonlinear topological pumping of solitons in spinor systems by utilizing the relative motion degrees of freedom between different spin components.\end{abstract}

\maketitle

\section{Introduction} 

\begin{figure}[tb]
\centering
\includegraphics[width=0.9\columnwidth]{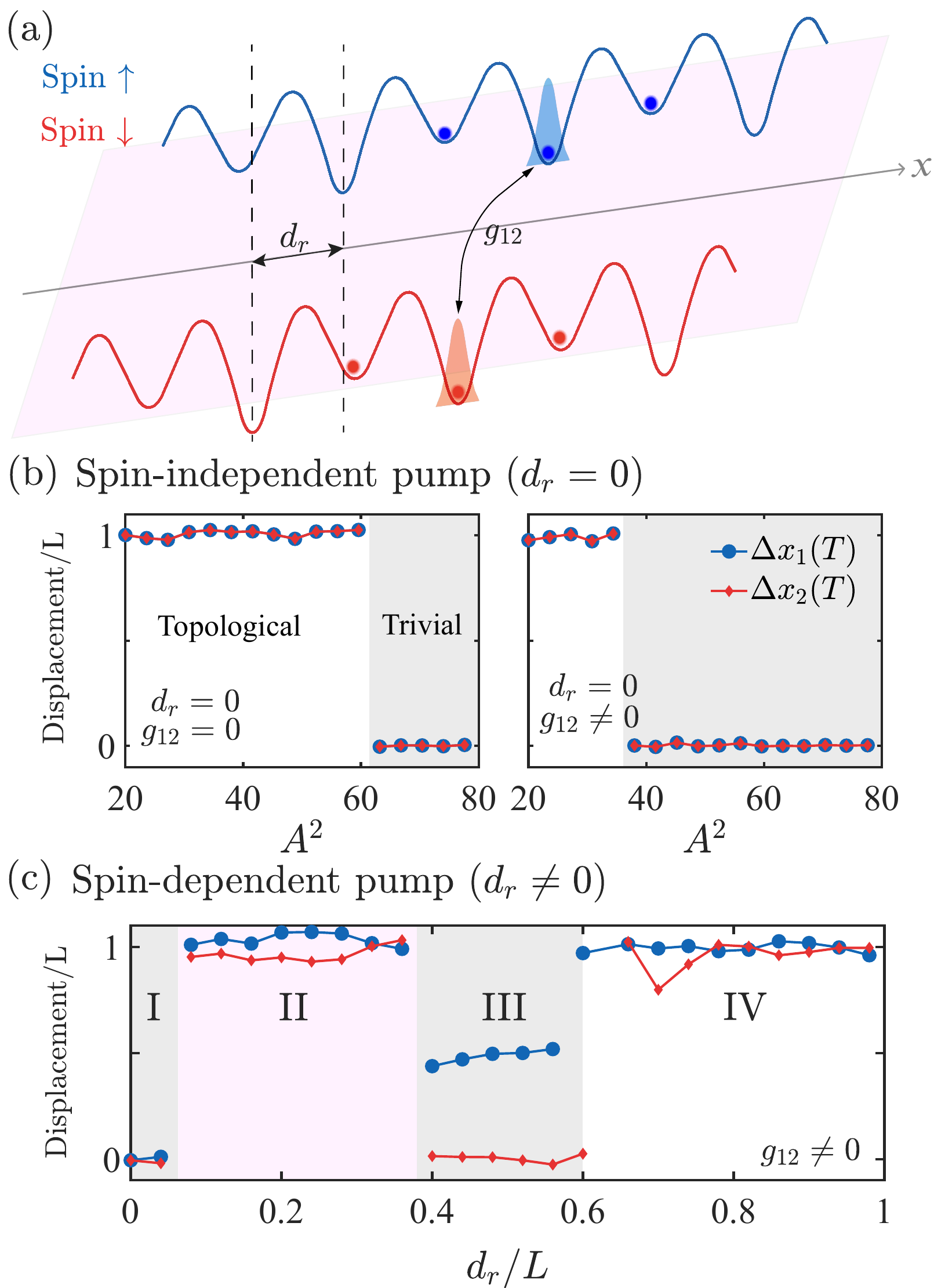}
\caption{Spin-dependent Thouless pumping of vector solitons in a quasi-1D two-component BEC. (a) System schematic. A quasi-1D two-component BEC is loaded in a spin-dependent superlattice in x-direction. The spinor BEC exhibits attractive intra- and inter-component interactions, characterized by the coupling constants $g$ and $g_{12}$, respectively. The spin-up and spin-down superlattices are of the same Thouless-pump-type [see Eq.~(\ref{super})] but are displaced by a distance $d_r$. (b) Phase diagram for spin-independent pumps ($d_r=0$). The mean displacements $\Delta x_j(T)$ ($j=1,2$) of the two components of the vector soliton at the end of one pump period $T$ are shown as a function of the nonlinearity strength $A^2$ for $g_{12}=0$ (left) and $g_{12}=0.6$ (right). (c) Phase diagram for spin-dependent pumps ($d_r\neq 0$). The $\Delta x_j(T)$ is shown as a function of $d_r$ for $g_{12}=0.6$ and $A^2=40$, displaying four distinct regimes I-IV. In both (b)-(c), we have $g=1$, and the superlattice potential with $V_s=V_l=25$ and $\nu=0.1$.}\label{Fig:Fig1}
\end{figure}

At present, there are significant interests in nonlinear topological pumping~\cite{Smirnova2020,Citro2023} in the context of synthetic materials~\cite{Leykam2016,Tangpanitanon2016,Bisianov2019,Ivanov2020,Tianxiang2021} and interacting quantum systems~\cite{Berg2011,Solnyshkov2017,Cuadra2020,Viebahn2024}. In particular, when nonlinear excitations such as solitons are subjected to a Thouless pump~\cite{Niu1984}, their motion can be quantized~\cite{Jurgensen2021,Mostaan2022,Qidong2022,Fu2022,Jurgensen2022,Tuloup2023,Jurgensen2023,Cao2024,TaoarXiv2024}, which is dictated by the topology of the underlying band structures. Such quantization has been shown to break down as nonlinearity becomes sufficiently strong~\cite{Jurgensen2021,Qidong2022,Fu2022,Tuloup2023,Cao2024}. The motivation behind the interests in soliton pumping is twofold. First, the soliton is associated with a non-uniform occupation of a Bloch band, violating the standard condition for quantized displacement in traditional, linear Thouless pumps~\cite{Kraus2012,Wang2013,Grusdt2014,Zeng2016,Lohse2016,Nakajima2016,Lu2016,Yongguan2016,Li2017,Ke2017,Taddia2017,Hayward2018,Wenchao2018,Nakagawa2018,Stenzel2019,Haug2019,Unanyan2020,Greschner2020,Chen2020,Cerjan2020,Nakajima2021,Minguzzi2022}. Second, nonlinear Thouless pumping provides new avenues for controlling the directional movement of matter waves, opening up new opportunities in quantum information processing~\cite{Smirnova2020,Citro2023}. So far, soliton pumping has been mainly studied in scalar nonlinear systems. 

Beyond scalar solitons, topological transport of vector solitons are of fundamental interest, as the spin degree of freedom~\cite{Ho2011,Li2012,Li2014,Chen2016,Wu2018} potentially enriches the interplay between the nonlinearity and topology~\cite{Berg2011,Nakagawa2018,Stenzel2019}. Recently, some studies in this direction have been carried out~\cite{Lyu2024}, showing transport phenomena without analogues in the counterpart of scalar solitons. However, these works have primarily considered spin-independent pumps. Meanwhile, substantial advancements in spin-dependent lattice techniques using ultracold atoms~\cite{Goldman2016} offer new opportunities for exploring the interplay between spin, nonlinearity, and topology in topological pumps. 

In this work, we explore Thouless pumping of vector solitons in a quasi-one-dimensional (quasi-1D) two-component Bose-Einstein condensate (BEC) confined within \textit{spin-dependent} optical superlattices [Fig.~\ref{Fig:Fig1}(a)]. The two-component (or spinor) BEC system exhibits both attractive intra- and inter-component interactions. The superlattice potentials for the spin-up and spin-down components are of the same Thouless-type but are displaced by a tunable distance $d_r$, constituting a spin-dependent Thouless pump. Consequently, the initial vector soliton state - a nonlinear excitation of the system - exhibits a relative distance $d_r$ between its two components. We show that the relative-motion degrees of freedom associated with this relative motion, characterized by $d_r$, serves as a novel control parameter for engineering nonlinear Thouless pumping.

Specifically, we numerically simulate the evolution of the vector soliton during the pumping cycles using the two-component Gross-Pitaevskii equation (GPE), with the emphasis on the roles played by the inter-component interaction and $d_r$. Our findings reveal that when $d_r=0$, the two soliton components can be either simultaneously pumped or arrested, depending on the effective interaction experienced by each component. Intriguingly, as we fix the interaction strength and increase $d_r$ to a nonzero value ($d_r\neq 0$), we find that an arrested vector soliton can re-enter the pumped regime, where both components undergo a quantized shift. However, as $d_r$ further increases, the solitons can get arrested again, whereas a revival of synchronized topological transport occurs with continuous increase of $d_r$. Using Lagrange variational method, we show that these intriguing phenomena are due to the significant impact of variations in $d_r$ on the inter-component interaction, as well as their effective modification of the lattice strength perceived by the center-of-mass of the vector soliton. Our work show that spinor systems in spin-dependent superlattices offer rich platforms for exploring and controlling nonlinear Thouless pumping.

The paper is organized as follows. In Sec.~\ref{Model}, we present detailed descriptions of our model system. In Sec.~\ref{Numerical}, by numerically solving GP equation, we investigate the nonlinear Thouless pumping of solitons with the emphasis
on the effects of inter-component interaction and $d_r$. In Sec.~\ref{Variational} we analytically derive the equations of motions for the vector soliton using Lagrangian variational approach. In Sec.~\ref{Conclude}, we discuss generality of our results, and summarize our work.

\begin{figure*}[t]
	\centering
	\includegraphics[width=0.85\textwidth]{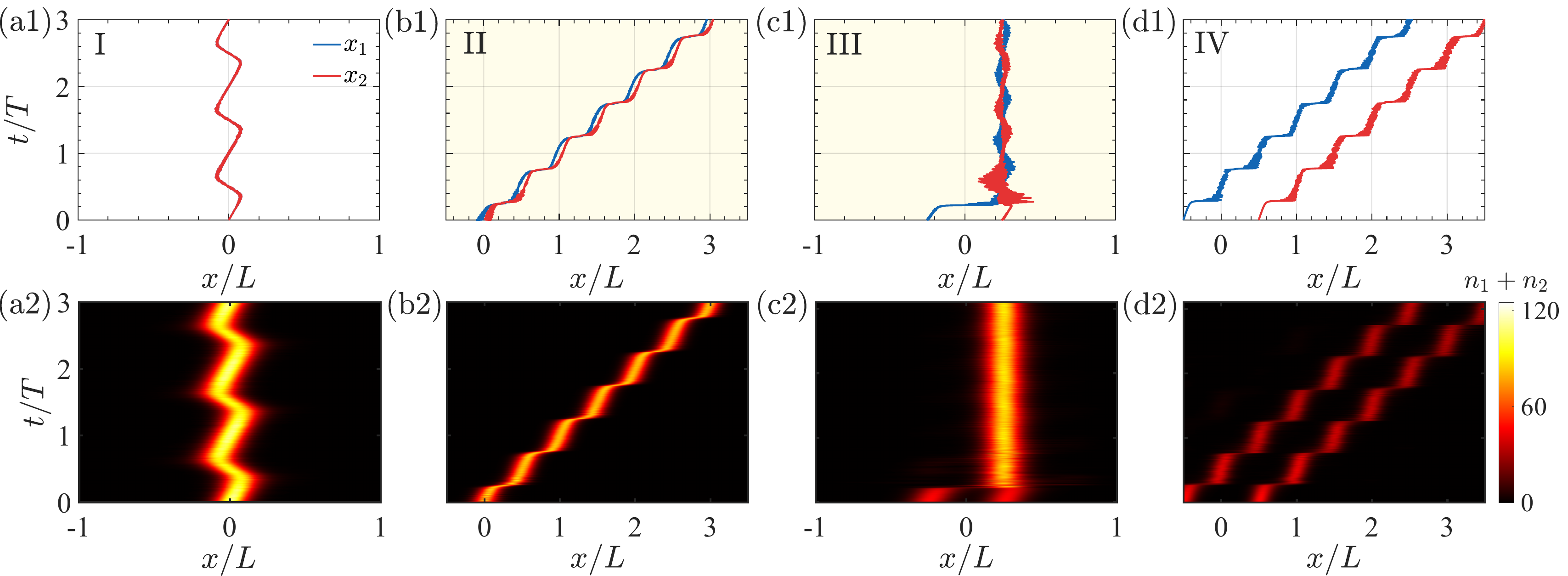}
	\caption{Numerical simulations of the dynamics of vector solitons in regimes I-IV in Fig.~\ref{Fig:Fig1}(c). Numerical results are obtained from solutions of GPE~(\ref{Eq:GPE}). The first row displays the mean displacement $x_j$ ($j=1,2$) of the vector soliton, and the second row displays the corresponding density distributions in space and time. Results are shown for (a1)-(a2) $d_r=0$, (b1)-(b2) $d_r=0.2$, (c1)-(c2) $d_r=0.5$, and (d1)-(d2) $d_r=1$. In all panels, we use $g=1$, $g_{12}=0.6$, $V_l=V_s=25$, $\nu=0.1$, and we consider the initial vector soliton state with the width $l=0.4$ and the intensity $A^2=40$.}\label{Fig:Fig2}
\end{figure*}

\section{Model System}\label{Model} 
In this work, we are interested in a two-component BEC trapped in a strongly anisotropic lattice 
potential: in the $x$ direction, the BEC is trapped in a spin-dependent superlattice $V_{\text p}(x,t)$ as illustrated in Fig.~\ref{Fig:Fig1}(a), while in 
the $y$ and $z$ directions, the strength of the transverse confinement with the large trap frequency $\omega_\perp$ is strong enough to freeze the atomic motion in these directions, allowing the atomic tunneling only in the $x$ direction. 
At the mean-field level, the energy functional~\cite{Ho2011,Li2012,Li2014,Chen2016,Wu2018} of our model system can be written as 
\begin{eqnarray}
H&=&\int d{\bf{r}}\Psi^\dagger({\bf{r}},t)\left(H_{0x}+H_{0\perp}\right) \Psi({\bf{r}},t)\nonumber\\
&+&\frac{1}{2}\int d{\bf{r}} \left[\left(g+g_{12}\right)n({\bf{r}},t)+\left(g-g_{12}\right)s_z({\bf{r}},t)\right], \label{Eq:H}
\end{eqnarray}
with $\Psi=\left[\Psi_1(x,t), \Psi_2(x,t)\right]^T$ being the wavefunction for the two-component BEC with the normalization of $\int d{\bf r}(|\Psi_1|^2+|\Psi_2|^2)=N$ and $N$ being the total number of atoms. The  $H_{0x}$ and $H_{0\perp}$ in Eq.~(\ref{Eq:GPE}) are referred as to the single-particle energy along the transverse and $x$ direction.
In more details, the $H_{0\perp}$ reads as $H_{0\perp}=(\hat{p}^2_y+\hat{p}^2_z)/2m+\frac{m}{2}\omega_{\perp}^2(y^2+z^2)$ and the  $H_{0x}$
contains the kinetic energy and a spin-dependent superlattice 
\begin{equation}
H_{0x}=\frac{\hat{p}^2_x}{2m}+V_{\text{p}}\left(x+\frac{d_{r}}{2}\sigma_z,t\right) .\label{SuperOL}
\end{equation}
Specifically, as shown in Eq. (\ref{SuperOL}), each spinor component is subjected to a superlattice potential of the same form, $
V_{\text{p}}(x)/E_R=-V_{\text{s}}\cos^2\left(\frac{2\pi x}{d}\right)-V_\text{l}\cos^2\left(\frac{\pi x}{d}-\Omega t\right)$ with the recoil energy $E_R=\hbar^2\pi^2/(2m d^2)$, but the spin-up and the spin-down superlattices are displaced from each other by a distance $d_r$. When $d_r=0$, the lattice configuration reduces to spin-independent Thouless-type pumps as experimentally realized in quantum gases~\cite{Lohse2016,Nakajima2016}, which consists of a primary lattice with the spatial period $d/2$ and lattice strength $V_{\text{s}}$, and a superimposed second lattice with the spatial period $d$ and the strength $V_l$; the relative phase between the two lattices varies in time at the rate $\Omega$. 

The $H_{\text{int}}$ in  Eq.~(\ref{Eq:GPE}) contains the intra-component and inter-component interactions between bosonic atoms, written 
in terms of the total density $n({\bf{r}},t)=|\Psi_1({\bf{r}},t)|^2+|\Psi_2({\bf{r}},t)|^2$ and the spin-density $s_z({\bf{r}},t)=|\Psi_1({\bf{r}},t)|^2-|\Psi_2({\bf{r}},t)|^2$. The $g=-4\pi \hbar^2 |a|/m$ and $g_{12}=-4\pi \hbar^2 |a_{12}|/m$ are the intra-component and inter-component coupling constant, with $a$ and $a_{12}$ being the s-wave scattering length. In this work, we restrict ourselves to the case $a<0$ and $a_{12}<0$.

Due to the strong transverse confinement, the BEC considered in this work is frozen to the ground state of the harmonic trap along these directions.
This allows a decomposition of the condensate wave function of the form $\Psi({\bf{r}},t)=\Psi(x,t)\phi(y,z)$ with $\phi(y,z)=1/(\sqrt{\pi}a_\perp)e^{-(y^2+z^2)/2a_{\perp}^2}$, so that after integrating out the transverse degrees of freedom, the considered
system is effectively treated as a quasi-1D two-component BEC described by an effective one-dimensional energy functional. Consequently, after rescaling the parameters
$x\rightarrow x/d$, $t \rightarrow 2E_Rt/\pi^2\hbar$, $\nu\rightarrow \Omega\hbar\pi^2/2E_R $, $N\rightarrow N/d$, $g\rightarrow -g\pi^2/4E_Ra^2_\perp$ and $g_{12}\rightarrow -g_{12}\pi^2/4E_Ra^2_\perp$, and following standard procedures, 
we obtain the dimensionless GPE for the quasi-1D BEC can be derived as
\begin{eqnarray}
	\!\!\!\!	i\frac{\partial \Psi_1}{\partial t}&\!\!=\!\!&\left[-\frac{1}{2}\frac{\partial^2}{\partial x^2}\!\!+\!\!V_\textrm{p}(x+\!\!\frac{d_r}{2},t)\!\!-\!\!g|\Psi_1|^2\!\!-\!\!g_{12}|\Psi_{2}|^2\right]\Psi_1,\nonumber\\
	\!\!\!\!	i\frac{\partial \Psi_2}{\partial t}&\!\!=\!\!&\left[-\frac{1}{2}\frac{\partial^2}{\partial x^2}\!\!+\!\!V_\textrm{p}(x-\!\!\frac{d_r}{2},t)\!\!-\!\!g_{12}|\Psi_{1}|^2\!\!-\!\!g|\Psi_2|^2\right]\Psi_2, \label{Eq:GPE}
	\end{eqnarray}
where $\Psi_{1}(x,t)$ and $\Psi_2(x,t)$ are the rescaled wavefunctions for the spin-down and spin-up components, respectively, with $\int dx (|\Psi_1|^2+|\Psi_2|^2)=N$. The pumping potential now takes the form
	\begin{eqnarray}
	V_\textrm{p}(x,t)=-V_s\cos^2\left(2\pi x \right)-V_l\cos^2\left(\pi x-\nu t\right), \label{super}
\end{eqnarray}
with the spatial period $L=1$ and time period $T=\pi/\nu$, and $\nu\ll 1$ under adiabatic condition. 

\section{Numerical Study}\label{Numerical}

We first numerically study the transport of the vector-soliton based on solutions of GPE~(\ref{Eq:GPE}), aiming at how its behavior is affected by the inter-component interaction characterized by $g_{12}$ and by the lattice displacement $d_r$. 

In our numerical simulations of the GPE, the system is initialized at time $t=0$ in a vector soliton state with $\Psi_j(x,0)=Ae^{-(x-x_j^0)^2/l^2}\varphi(x-x_j^0,0)$ for $j=1, 2$. Here, $A$ is the initial peak amplitude, $l$ is the initial width of the envelope. The $\varphi(x-x_j^0,0)$, with  $x_j^0=(-1)^jd_r/2$, is the lowest Bloch state at $k=0$ associated with the linear part of the Hamiltonian in Eq.~(\ref{Eq:GPE}). That is, the spin-up and spin-down soliton components are initially separated by a distance $d_r$ in the $x$-direction. Starting from this initial condition, we numerically solve the GPE using split-step fast Fourier algorithm. Details of our numerical scheme can be found in Appendix~\ref{AppendixA}. 

The quantities of interest are the center-of-mass positions of the components $j=1,2$, given by 
\begin{align}
x_j=\frac{\int x|\Psi_j|^2dx}{\int |\Psi_j|^2dx}.
\end{align}
 The topological property is characterized by the mean displacement
\begin{equation}
\Delta x_j=x_j-x_j^0.  \label{deltax}
\end{equation}
For later analysis, we also introduce the center-of-mass position $x_c$ of the vector soliton and the relative displacement $x_r$ between its two components, i.e., 
\begin{equation}
x_c=(x_1+x_2)/2, \hspace{2mm}x_r=(x_1-x_2)/2. \label{relative}
\end{equation}

We will analyze two scenarios: (i) Spin-independent pumps (i.e., $d_r=0$), where the superlattice potentials acting on the spin-up and spin-down components coincide; (ii) Spin-dependent pumps (i.e., $d_r\neq 0$), where the pumps are spin dependent, with the superlattices in different components being displaced from each other along $x$-direction. 

Without loss of generality, we illustrate our study with the pumping potential in Eq.~(\ref{super}) with $V_s=V_l=25$. In this case, as detailed in Appendix~\ref{AppendixB}, the lowest three linear Bloch bands in the space-time domain are characterized by the Chern numbers $C_1=1$, $C_2=-1$, $C_3=1$, respectively. We also consider a slow modulation rate $\nu=0.1$ that fulfills the adiabatic condition. 

\subsection{Spin-independent Pumps: $d_r=0$}

When $d_r=0$, the pumps become spin-independent. Therefore, initially at $t=0$, the vector soliton state as a nonlinear eigensolution of the GPE~(\ref{Eq:GPE}) contains two components whose positions coincides with each other. 

Without inter-component interaction, i.e., $g_{12}=0$, the two components are independent of each other, and the problem is reduced to the pumping of a scalar soliton extensively studied in previous works~\cite{Jurgensen2021,Mostaan2022,Qidong2022,Jurgensen2022,Fu2022,Jurgensen2023,Cao2024}. There, a soliton can be pumped or arrested, depending on the nonlinear strength characterized by $gA^2$. Specifically, for $gA^2$ below some threshold, the soliton undergoes a quantized transport; for $gA^2$ above the threshold, the soliton is arrested near its initial position. As illustrated in the left panel of Fig.~\ref{Fig:Fig1}(b), where $g=1$ and $g_{12}=0$, our numerical results of $\Delta x_j(T)$ after one pump period $T$ reproduce the scalar BEC case in Ref.~\cite{Qidong2022}. Specifically, for $gA^2<61$, each component undergoes a quantized shift with $\Delta x_j=1$, but for stronger nonlinearities $gA^2>61$, both components are trapped near their initial position, with $\Delta x_j=0$ . 

The situation becomes different when the inter-component interaction is turned on, i.e., $g_{12}\neq 0$. In this case, the soliton in one component experiences an additional attractive interaction from the other component. Therefore, the effective nonlinear interaction seen by one component is enhanced to $\sim (g+g_{12})A^2$. As illustrated in the right panel of Fig.~{\ref{Fig:Fig1}}(b), where $g=1$ and $g_{12}=0.6$, both components are simultaneously pumped or arrested, but the pumped-to-arrested transition is enhanced by the presence of attractive inter-component interaction $g_{12}\neq 0$ compared to the case with $g_{12}=0$. We can estimate the critical value of $A_c$ for the phase transition from $(g+g_{12})A_c^2\approx61$, yielding $A_c\sim 38$, which is consistent with the numerical results.

\begin{figure}[t]
\centering
\includegraphics[width=1\columnwidth]{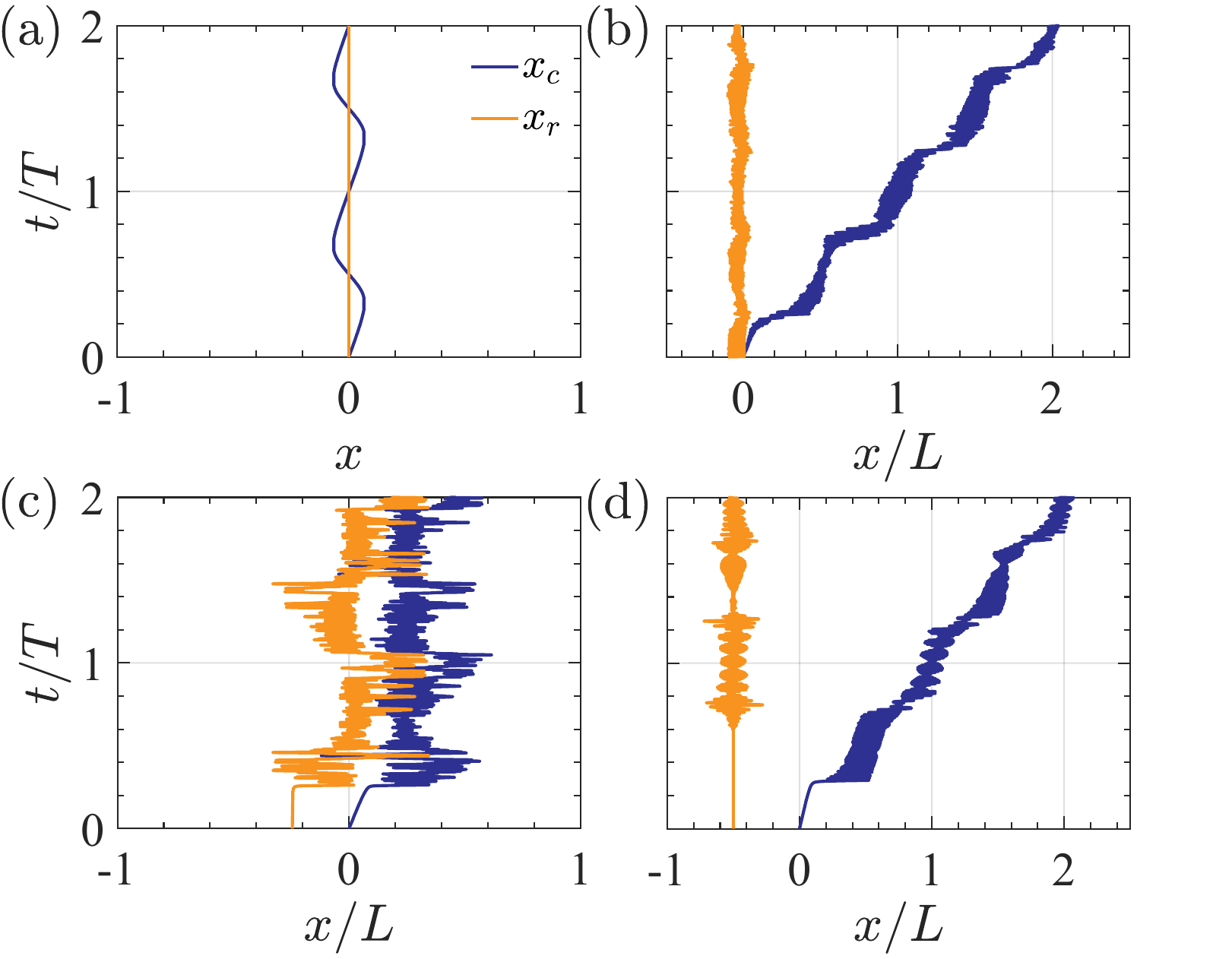}
\caption{Dynamics of the vector soliton obtained from the variational approach under various $d_r$. Results for the center-of-the-mass position $x_c=(x_1+x_2)/2$ of the vector soliton and the relative displacement $x_r=(x_1-x_2)/2$ of its spin components are obtained from solutions of Eq.~(\ref{Eq:EqOfMotion}) for (a) $d_r=0$, $\sigma_1(0)=\sigma_2(0)=0.1064$, (b) $d_r=0.2$, $\sigma_1(0)=\sigma_2(0)=0.1398$, (c) $d_r=0.49$, $\sigma_1(0)=\sigma_2(0)=0.1299$, and (d) $d_r=1$, $\sigma_1(0)=\sigma_2(0)=0.1296$. For other parameters, we take $A^2=23$, $V_s=V_l=25$, $g=1$, $g_{12}=0.6$ and $\nu=0.1$. In addition, we consider $\dot{x}_c(0)=\dot{x}_r(0)=\dot{\sigma}_1(0)=\dot{\sigma}_2(0)=0$. }\label{Fig:Fig3}
\end{figure}

\subsection{Spin-dependent Pumps: $d_r\neq 0$}

\begin{figure*}[t]
	\centering
	\includegraphics[width=0.85\textwidth]{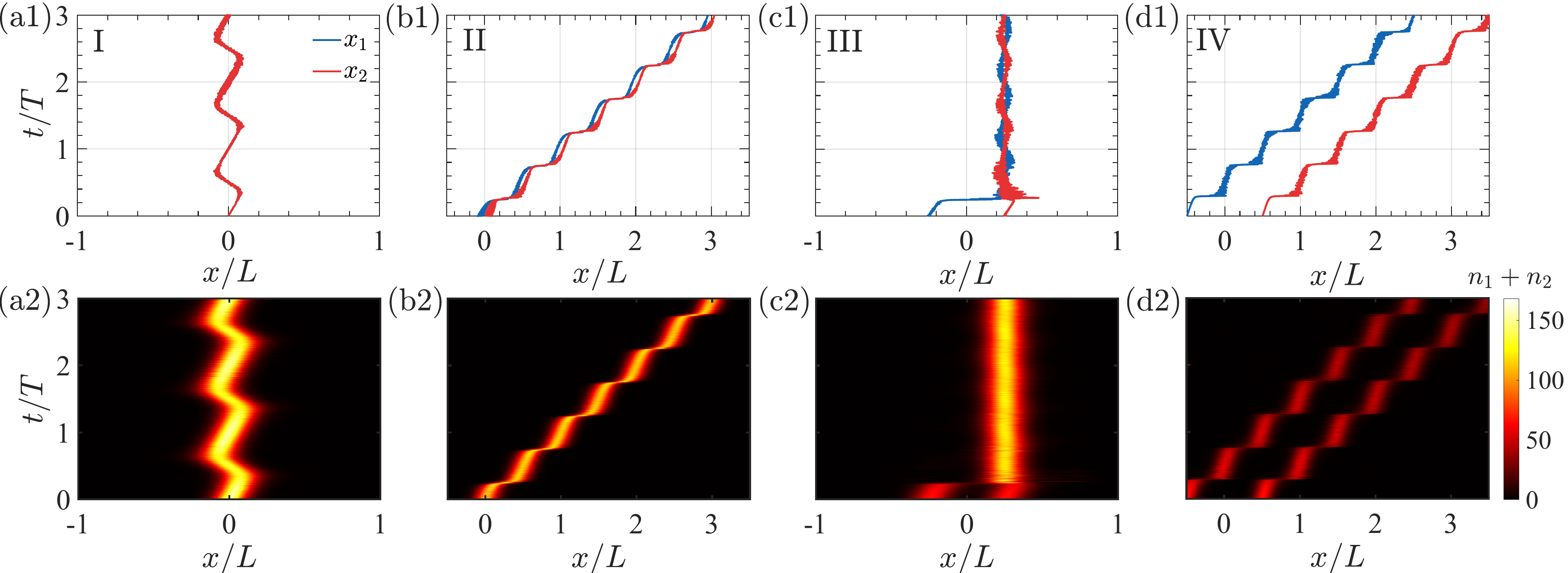}
	\caption{Transport dynamics of vector solitons for nonlinearity parameters $g=0.8$, $g_{12}=0.4$ and $A^2=55$. Numerical results are obtained from solutions of GPE (\ref{Eq:GPE}). The first row displays $x_j$ ($j=1,2$) of the vector soliton, and the second row displays the corresponding density distributions in space and time. Results are shown for (a1)-(a2) $d_r=0$, (b1)-(b2) $d_r=0.2$, (c1)-(c2) $d_r=0.5$, and (d1)-(d2) $d_r=1$. In all panels, $l=0.4$ and the pumping potentials are the same as in Fig.~\ref{Fig:Fig2}.}\label{Fig:Fig4}
\end{figure*}

Compared with the intra-component interaction, the inter-component attractive interaction not only relies on $g_{12}A^2$, it also depends on the separation $|x_1(t)-x_2(t)|$ between the spin-up and spin-down solitons at time $t$. Therefore, the inter-lattice displacement $d_r$ is expected to provide an additional control for the effective nonlinearity. To illustrate the essential physics, let us fix other nonlinearity parameters to be $g=1$, $g_{12}=0.6$ and $A^2=40$, so that when $d_r=0$, the vector soliton is in the trapped phase [see Fig.~{\ref{Fig:Fig1}}(b), right panel]. We then vary $d_r$ and numerically solve the corresponding GPE (\ref{Eq:GPE}). In Fig.~\ref{Fig:Fig1}(c), we present the resulting $\Delta x_j(T)$ ($j=1,2$) at the end of one pump cycle $T$ as a function of $d_r$, showing a rich phase diagram. Four distinct phases can be identified: (1) Phase I: for small $d_r$, both solitons remain arrested; (2) Phase II: when $d_r$ increases, the solitons become topologically pumped; (3) Phase III: when $d_r$ further increases, one soliton becomes arrested again, and the other soliton component exhibits fractional shift at $T$; (4) Phase IV: when $d_r$ becomes sufficiently large, both solitons again synchronize and exhibit a quantized shift.  

Typical dynamical behaviors of the vector soliton in phases I-IV are illustrated in Figs.~\ref{Fig:Fig2}(a)-(d), which can be qualitatively understood as follows. In phase I [Fig.~\ref{Fig:Fig2}(a)], where the initial solitonic separation is much smaller than its width (i.e., $d_r\ll l$), the overlap between the wavefunctions of the spin-up and spin-down components is substantial at $t=0$, so that the two solitons are essentially bounded together during the pumping process. In this scenario, the physics is similar as the $d_r=0$ case: it can be understood using the center-of-the-mass degrees of freedom of the vector soliton, while the relative motion can be ignored. In the opposite limit $d_r\gg l$ in phase IV, the overlap between the two components is small and the inter-component interaction is ignorable. In this case, the dynamics of the vector soliton can be understood in terms of two noninteracting scalar solitons. As shown in Fig.~(\ref{Fig:Fig2})(d), for $g=1$ and $A^2=40$, each soliton undergo a quantized transport, and their relative distance is maintained at the initial value during the pump.

The dynamical behaviors in phases II and III [Figs.~\ref{Fig:Fig2}(b)-(c)] reveal an interesting interplay between the relative motion between the two soliton components and the topological transport. In phase II [Fig.~\ref{Fig:Fig3}(b)], where $d_r\lesssim l$, the attractive interaction between the two solitons is suppressed but remains finite. This reduction in the inter-component interaction leads to a quantized shift of both components, accompanied by small temporal oscillations in their relative distance. Interestingly, in phase III [Fig.~\ref{Fig:Fig2}(c)] where $d_r\gtrsim l$, we observe that one soliton (blue curve) is pulled toward the other (red curve) during the initial stage of the pump, and when they almost coincide, both become arrested. This behavior can be intuitively understood by noting that the pumping potential (\ref{super}) generates identical rightward force on both solitons. However, the attractive inter-component interaction has opposite effects on the two components: it increases the rightward force on one soliton, facilitating its transport along the pump direction, while it decreases the force on the other soliton, slowing its rightward motion. As a result, the two solitons are gradually pulled together over time, leading to an increase in the inter-component interaction. Eventually, the total effective nonlinearity experienced by each component is enhanced to arrest the motion of the soliton [Fig.~\ref{Fig:Fig2}(c)].

\section{Variational approach}\label{Variational}

To further understand the intriguing behaviors of the vector soliton under the above spin-dependent pumps, in this section, we analytically study the motion of the vector soliton using the Lagrangian variational approach.

To begin with, we first introduce a transformation $\Psi_j(x,t)=\phi_j(x,t)e^{i(V_s+V_l)t/2}$ for the $j$th component, and recast Eq.~(\ref{Eq:GPE}) into the form 
\begin{eqnarray}
\!\!\!\!\!i\frac{\partial \phi_1}{\partial t}\!\!&=&\!\!\left[\!-\frac{1}{2}\frac{\partial^2}{\partial x^2}+\!\mathcal{V}_\textrm{p}(x+\frac{d_r}{2},t)-\!\!g|\phi_1|^2\!-\!g_{12}|\phi_{2}|^2\right]\!\phi_1,\nonumber\\
\!\!\!\!\!i\frac{\partial \phi_2}{\partial t}\!\!&=&\!\!\left[\!-\frac{1}{2}\frac{\partial^2}{\partial x^2}+\!\mathcal{V}_\textrm{p}(x-\frac{d_r}{2},t)-\!\!g_{12}|\phi_{1}|^2\!-\!g|\phi_2|^2\right]\!\phi_2,\label{Eq:ScalingGPE}
\end{eqnarray}
with the transformed potential $\mathcal{V}_\textrm{p}(x,t)=-\frac{V_s}{2}\cos\left(4\pi x\right)-\frac{V_l}{2}\cos(2\pi x-2\nu t)$. We assume the variational ansatz for the component $j=1,2$ as
\begin{eqnarray}
	\phi_j(x)=\frac{A}{\sqrt{\sqrt{4\pi}\sigma_j}}e^{-\frac{1}{2}\big(\frac{1}{\sigma_j^2}-i\frac{\dot{\sigma}_j}{\sigma_j}\big)(x-x_j)^2+i\dot{x}_j(x-x_j)},\label{Eq:ansatz}
\end{eqnarray}
where the variational parameters $\sigma_j(t)$ and $x_j(t)$, which describe the width and center-of-the-mass position of the $j$th component. Moreover, $\dot{x}_j$ and $\dot{\sigma}_j$ label the velocity and the change of width in time for the soliton in component $j$, respectively.

The Lagrangian $\mathbb{L}=\int_{-\infty}^{+\infty}\mathcal{L}dx$ corresponding to Eq.~(\ref{Eq:ScalingGPE}) can be derived following the standard procedures~\cite{Victor1996,Navarro2009,Xu2019}. We obtain 
\begin{eqnarray}
	\mathcal{L}&=&\sum_{j=1,2}\left[\frac{i}{2}\left(\phi_j^\star\frac{\partial}{\partial t}\phi_j-\phi_j\frac{\partial}{\partial t}\phi_j^\star\right)-\frac{1}{2}\left|\frac{\partial\phi_j}{\partial x}\right|^2+\frac{1}{2}g|\phi_j|^4\right]\nonumber\\
	&&-\mathcal{V}_p(x+\frac{d_r}{2},t)|\phi_1|^2-\mathcal{V}_p(x-\frac{d_r}{2},t)|\phi_2|^2\nonumber\\
	&&+g_{12}|\phi_1|^2|\phi _2|^2. \label{Lag}
\end{eqnarray}
Inserting the variational ansatz~(\ref{Eq:ansatz}) into Eq.~(\ref{Lag}) and switching to new variables $x_c$ and $x_r$ in Eq.~(\ref{relative}), after some algebra, we obtain the Lagrangian in the form $\mathbb{L}=T-V_\textrm{ex}$. Here, the kinetic energy is $T=\frac{A^2}{2}{\dot{x}_c}^2+\frac{A^2}{2}{\dot{x}_r}^2-\frac{\sigma_1A^2}{8}\ddot{\sigma}_1-\frac{\sigma_2A^2}{8}\ddot{\sigma}_2$, and the effective potential energy is given by
\begin{widetext}
	\begin{eqnarray}
		V_\textrm{ex}&=&\sum_{j=1,2}\left(\frac{A^2}{8\sigma_j^2}-\frac{gA^4}{8\sqrt{2\pi}\sigma_j}\right)-\frac{g_{12}A^4}{4\sqrt{\pi(\sigma_1^2+\sigma_2^2)}} e^{-4x_r^2/(\sigma_1^2+\sigma_2^2)}\nonumber\\
		\nonumber\\
		&-&\frac{A^2}{4}\left\{V_1 e^{-4\pi^2\sigma_1^2}\cos\left[4\pi\left(x_c+x_r+\frac{d_r}{2}\right)\right]+V_2e^{-\pi^2\sigma_1^2}\cos\left[2\pi\left(x_c+x_r+\frac{d_r}{2}\right)-2\nu t\right]\right.\nonumber\\
		&+&\left.V_1e^{-4\pi^2\sigma_2^2}\cos\left[4\pi\left(x_c-x_r-\frac{d_r}{2}\right)\right]+V_2e^{-\pi^2\sigma_2^2}\cos\left[2\pi\left(x_c-x_r-\frac{d_r}{2}\right)-2\nu t\right]\right\}. \label{Vex}
	\end{eqnarray}
\end{widetext}

The set of Euler-Lagrange equations for the four variational parameters are given by $\frac{\partial \mathbb{L}}{\partial y_i}-\frac{d}{dt}(\frac{\partial \mathbb{L}}{\partial \dot{y}_i})=0$, with $y_{i}=x_c, x_r, \sigma_1,\sigma_2$. After some tedious but straightforward calculations (see Appendix~\ref{AppendixC}), we can associate the time evolution of $y_i$ with the classical motion of a fictitious particle with an effective mass in the effective potential $V_{\text{ex}}$~(\ref{Vex}),
and derive the following equations of motion for $y_i$ as [see Eq.~(\ref{AppEqOfMotion}) in Appendix \ref{AppendixC}]
\begin{eqnarray}
m^i_{\text{eff}} \frac{d^2{y_i}}{dt^2}&=&-\frac{\partial V_\textrm{ex}}{\partial y_i},\label{Eq:EqOfMotion}
\end{eqnarray}
with the effective mass being $m_{\text{eff}}=[A^2,A^2,A^2/8,A^2/8]$. 

Equations~(\ref{Eq:EqOfMotion}) are supplemented with the following initial conditions: $x_c(0)=0$, $x_r(0)=-d_r/2$, $\dot{x}_c(0)=0$, $\dot{x}_r(0)=0$, and $\dot{\sigma}_1(0)=0$, $\dot{\sigma}_2(0)=0$. With these choices, the values of $\sigma_1(0)$ and $\sigma_2(0)$ are determined by the requirement to satisfy the equations $\partial \mathbb{L}/\partial {\sigma_1}=d\left(\partial \mathbb{L}/{\partial \dot{\sigma}_1}\right)/dt$ and $\partial \mathbb{L}/\partial{\sigma_2}=d\left(\partial \mathbb{L}/{\partial \dot{\sigma}_2}\right)/dt$ at time $t=0$. For $g_{12}=0$, solutions of Eq.~(\ref{Eq:EqOfMotion}) fully recover the results of the single-soliton case. 

In Fig.~\ref{Fig:Fig3}, we show the results for the center-of-mass position $x_c(t)$ of the two-component BEC and the relative displacement $x_r(t)$ between the two soliton components for various $d_r$, by solving Eq.~(\ref{Eq:EqOfMotion}). We see four types of distinct transport behaviors, qualitatively agreeing with the numerical simulations based on GPE.

\begin{figure}[t]
	\centering
	\includegraphics[width=1\columnwidth]{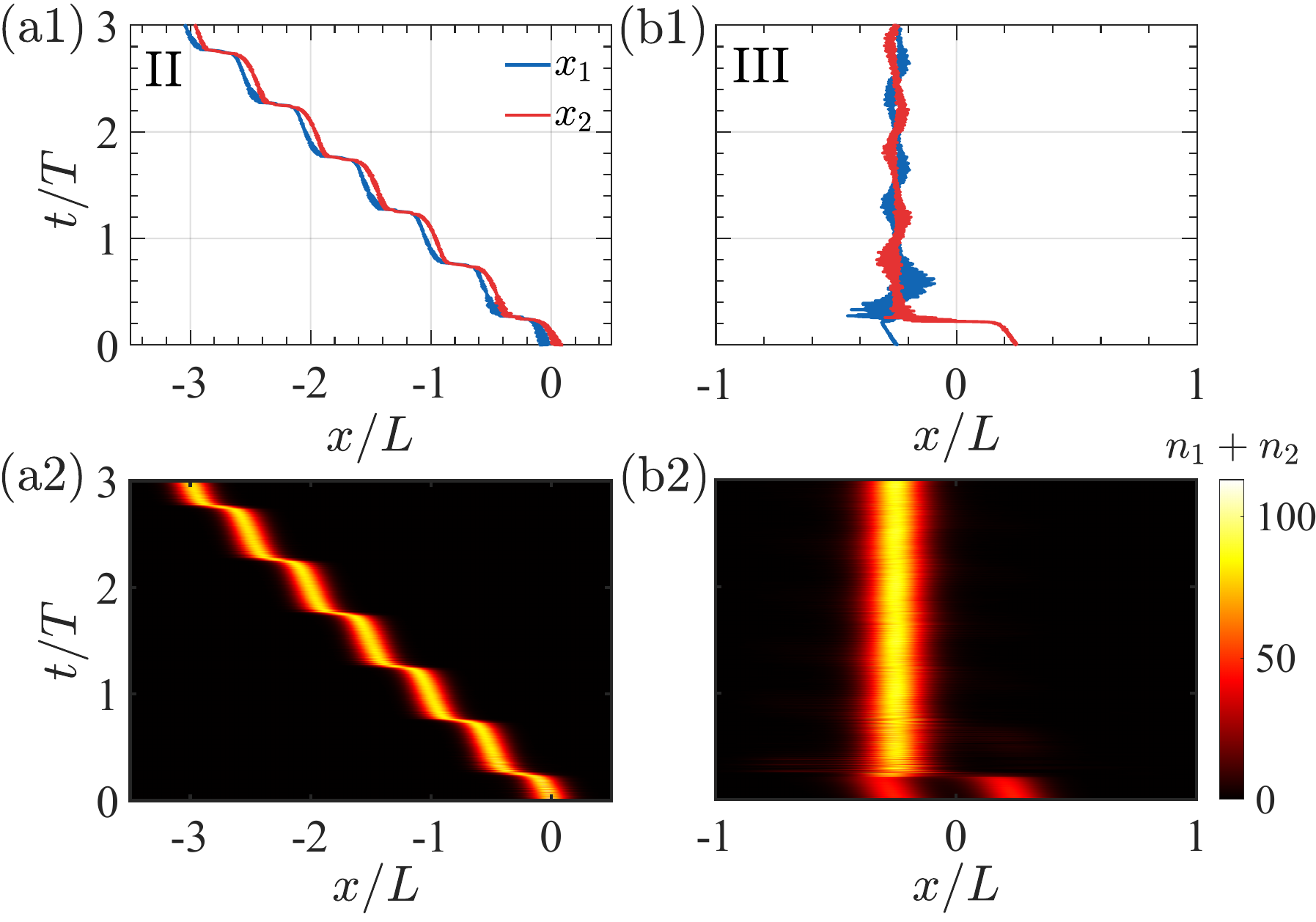}
	\caption{Numerical simulations of the vector-soliton dynamics under a leftward Thouless pumping. Results are obtained from numerical solutions of GPE (\ref{Eq:GPE}) with the superlattice potentials in the form ${V}_\textrm{p}(x,t)=-V_s\cos^2\left(2\pi x \right)-V_l\cos^2\left(\pi x+\nu t\right)$, shown for (a) $d_r=0.2$ and (b) $d_r=0.5$. Other parameters are the same as Fig.~\ref{Fig:Fig2}(c). }\label{Fig:Fig5}
\end{figure}

We remark that the effective potential in Eq.~(\ref{Vex}) allows us to qualitatively understand regarding how the relative motion of solitons controls the nonlinear topological phase transition of vector solitons. Specifically, the effective potential in Eq.~(\ref{Vex}) involves three contributions: (i) The first two terms in the bracket of the first line in Eq.~(\ref{Vex}) originate from the kinetic energy and intra-interaction energy, respectively. The competition of these two terms results in the existence of solitons for $g_{12}=0$. (ii) The third term in the first line comes from the inter-component interaction, which depends on both $g_{12}$ and the instantaneous relative distance $|x_r|$ between the two components. (iii) The second and third lines involve the spin-dependent pumps, which can be intuitive interpreted as follows: for the center-of-mass degrees of freedom $x_c$, it sees an effective superlattice whose strengths are modified by $d_r$. Therefore, Eq.~(\ref{Vex}) reveals the two aspects of the effects of the relative degrees of freedom: modulations of $d_r$ not only influence the distance-dependent inter-component interaction, but also effectively tune the superlattice potential seen by the center-of-the-mass of the vector soliton.

\section{Concluding Discussion}\label{Conclude}

Our results are generic under different choice of system parameters, as exemplified in Fig.~\ref{Fig:Fig4} for $A^2=55$, $g=0.8$ and $g_{12}=0.4$. In addition, the intriguing soliton dynamics seen in phases II and III, where the relative degrees of freedom plays an influential role, does not depend on the direction of the pump. For instance, in Fig.~\ref{Fig:Fig5}, a leftward Thouless pumping is considered, realized by two superlattices of the form ${V}_\textrm{p}(x,t)=-V_s\cos^2\left(2\pi x \right)-V_l\cos^2\left(\pi x+\nu t\right)$, which are displaced by $d_r$. The pumping dynamics of the vector soliton for $d_r=0.2$ [Fig.~\ref{Fig:Fig5}(a1)-(a2)] and $d_r=0.5$ [Fig.~\ref{Fig:Fig5}(b1)-(b2)] are shown. Still, we see similar dynamical behaviors as the case of rightward pumping in phases II and III [Fig.~(\ref{Fig:Fig2})]. 

In summary, we have studied the transport of vector solitons under spin-dependent Thouless pumps in form of spin-dependent superlattices shifted from each by a distance $d_r$. We show that modulations of $d_r$ leads to enriched nonlinear topological phase diagrams. This phenomenology can be understood because the relative motion of the vector soliton affects the inter-component interaction and also effectively modifies the external pump acting on the center-of-mass of the vector soliton.  
In summary, we have studied the transport of vector solitons under spin-dependent Thouless pumps in form of spin-dependent superlattices shifted from each by a distance $d_r$. Our results highlight the relative-motion degrees of freedom between different spin components offer an additional control knob of the nonlinear topological transport in spinor systems.

\section{Acknowledgement}
We thank Qidong Fu, Fangwei Ye, Qi Zhang, and Biao Wu for stimulating discussions. This work was supported by the National Natural Science Foundation of China (Nos. 12074344,12374246), the Zhejiang Provincial Natural Science Foundation (Grant Nos. LZ21A040001) and the key projects of the Natural Science Foundation of China (Grant No. 11835011).

\appendix

\section{Split-Step Fast Fourier Numerical Method}\label{AppendixA}

Following Ref.~\cite{Feit1979}, we present details on using the split-step fast Fourier algorithm to numerically solve the dimensionless GPE.~(\ref{Eq:GPE}) in the main text.

We first recast Eq.~(\ref{Eq:GPE}) into the following form
\begin{eqnarray}
	i\frac{\partial\Psi_j}{\partial t}=\left(\hat{O}_1+\hat{O}_2\right)\Psi_j,\label{AppendAGPE}
\end{eqnarray}
with 
\begin{eqnarray}
	\hat{O}_1&=&-\frac{1}{2}\frac{\partial^2}{\partial x^2},\\
	\hat{O}_2&=&V_\textrm{p}\left(x,t\right)-g|\Psi_j|^2-g_{12}|\Psi_{j'}|^2,\label{AppendAGPE3}
\end{eqnarray}
with $j'\neq j$. 

To numerically solve Eq.~(\ref{AppendAGPE}), we employ the second-order splitting scheme and write
\begin{eqnarray}
	\Psi(x, t+dt)=U_1U_2U_1\Psi(x,t),\label{AppendAGPET}
\end{eqnarray}
with $U_1=\exp(-iO_1dt/2)$ and $U_2=\exp(-iO_2dt)$. In our numerics, we choose a sufficiently small time step $dt=10^{-5}$.

Finally, we use the fast Fourier transform method to calculate each step of the time evolution in Eq.~(\ref{AppendAGPET}), and rewrite Eq. (\ref{AppendAGPET}) as
\begin{eqnarray}
	\!\!\!\!\!\!\!\!&&\Psi(x,t+dt)\nonumber\\
	\!\!\!\!\!\!\!\!&&=\mathcal{F}^{-1}\left[e^{-\frac{ik^2}{4}dt}\mathcal{F}\left(U_2\mathcal{F}^{-1}\left[e^{-\frac{ik^2}{4}dt}\mathcal{F}[\Psi(x,t)]\right]\right)\right],
\end{eqnarray}
where $\mathcal{F}$ and $\mathcal{F}^{-1}$ label the Fourier transform and inverse Fourier transform, respectively. In our numerical calculations, we choose $dx=2.5\times10^{-3}$ in the spatial discretization of GPE.

\bigskip

\section{Band structure and Chern number}\label{AppendixB}

\begin{figure*}[tb]
	\centering
	\includegraphics[width=1\textwidth]{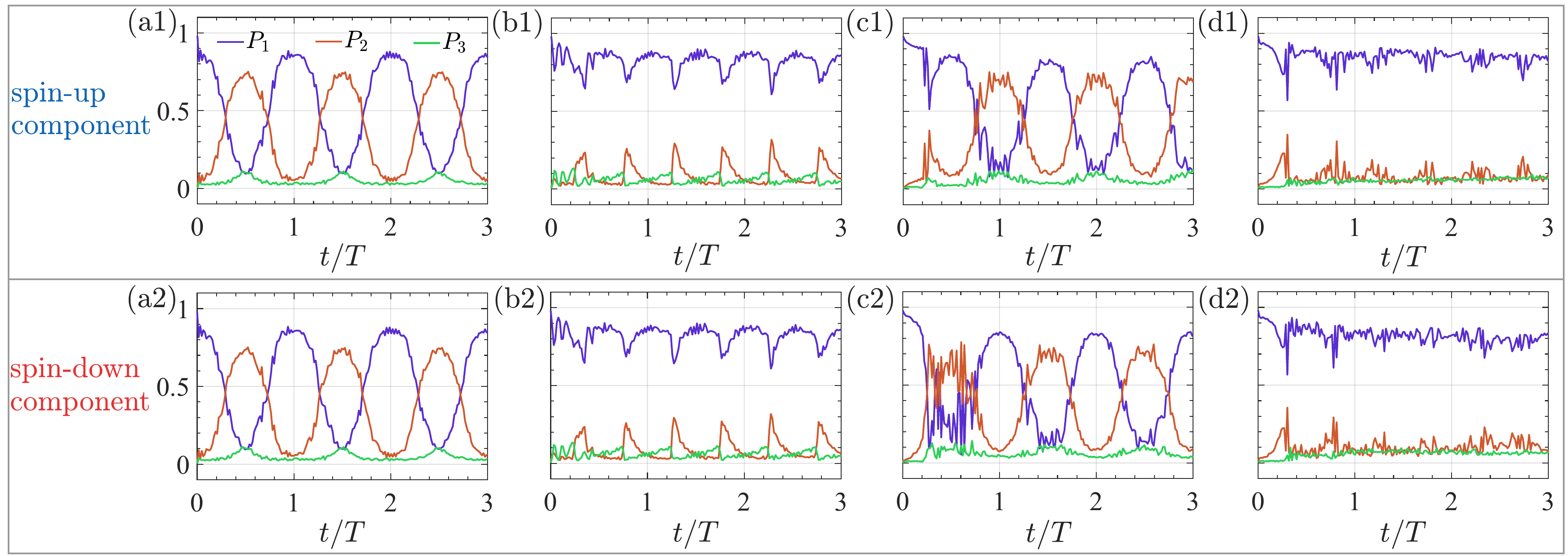}
	\caption{Numerical projections of the first component (in the first row) and the second component (in the second row) of the vector soliton state onto the linear Bloch bands. Occupations $P_\alpha$ ($\alpha=1,2,3$) are shown for the lowest three bands with Chern numbers $C=1,-1,1$.}\label{Fig:FigApp1}
\end{figure*}
In this section, we present details on (i) calculations of the Chern number of the Bloch bands associated with the linear Thouless pump~\cite{Di2010}, and (ii) projections of a soliton state onto these Bloch bands. 

First, we describe how to obtain the band structures of the linear Hamiltonian for our model system and determine the corresponding Chern numbers. Without interactions, Eq.~(\ref{Eq:GPE}) of the main text reduces to the scalar BEC case, where the wavefunction (of one component) satisfies
\begin{eqnarray}
	i\frac{\partial}{\partial t}\psi(x,t)=H_0(x,t)\psi(x,t),\label{Eq:LinearH}
\end{eqnarray}
with the Hamiltonian 
\begin{equation}
H_0=-\frac{1}{2}\frac{\partial^2}{\partial x^2}-V_s\cos^2\left(2\pi x \right)-V_l\cos^2\left(\pi x-\nu t\right),\label{H0xt}
\end{equation}
which has spatial periodicity $L=1$. Using the Bloch's theorem, we write $\psi(x,t)=e^{-ikx}u(x,t)$ with $u(x,t)=u(x+1,t)$, which can be expanded as
\begin{eqnarray}\label{Eq:BlochExpand}
	\psi(x,t)=e^{-ikx}\sum_{n=-M}^{+M}c_n(t)e^{i2n\pi x}.
\end{eqnarray}
Here, $c_n$ is the expansion coefficient,  and $M>0$ is a positive integer. Inserting Eq.~(\ref{Eq:BlochExpand}) to Eq.~(\ref{Eq:LinearH}), we obtain
\begin{eqnarray}
	i\frac{\partial}{\partial t} {\bf{c}} =H_0(k,t) {\bf{c}},
\end{eqnarray}
with ${\bf{c}}=(\dots,c_{-1},c_0,c_1,\dots)^T$. 

In numerical calculations, the expansion is typically truncated at $n_{\textrm{max}}=\pm M$, yielding a truncated matrix $H_0(k,t)$ with the dimension $(2M+1)\times(2M+1)$. The matrix form of $H_0$ is then given by
\begin{widetext}
	\begin{equation}
		H_0(k,t)=\frac{1}{4}\left(\begin{array}{cccccccccc}
			\ddots&&&&\\
			\cdots&h_{-2}&-{V_2}\Theta(t)&-{V_1}&&\\
			\cdots&-{V_2}\Theta^\star(t)&h_{-1}&-{V_2}\Theta(t)&-{V_1}\\
			&-{V_1}&-{V_2}\Theta^\star(t)&h_0&-{V_2}\Theta(t)&-{V_1}\\
			&&-{V_1}&-{V_2}\Theta^\star(t)&h_1&-{V_2}\Theta(t)&\cdots\\
			&&&-{V_1}&-{V_2}\Theta^\star(t)&h_2&\cdots\\
			&&&&&&\ddots&
		\end{array}\right)_{(2M+1)\times(2M+1)}.
	\end{equation}
\end{widetext}
Here, $h_n=2(k-n\Omega)^2-2V_1-2V_2$ for $n=-M,\cdots,M$, and $\Theta(t)=\exp(2i\nu t)$. In the detailed calculations, we have double-checked that $M=10$ is sufficient for calculating the Chern numbers of the band structures. 

We then diagonalize $H_0(k,t)$  to obtain the adiabatic right eigenstate $|\varphi_\alpha(k,t)\rangle$ and its corresponding eigenvalue $\epsilon_\alpha(k,t)$ ($\alpha=1,\dots,2M+1$). The Chern number $C_\alpha$ associated with the $\alpha$th band is then given by
\begin{eqnarray}
\!\!\!C_\alpha=\frac{i}{2\pi}\!\int_0^T\!\!dt\int_0^{\frac{2\pi}{L}}\!\!dk\left[\langle\partial_t\varphi_\alpha|\partial_k\varphi_\alpha\rangle\!-\!\langle\partial_k\varphi_\alpha|\partial_t \varphi_\alpha\rangle\right].
\end{eqnarray}
For the pumping potentials of the form in Eq.~(\ref{SuperOL}) with $V_l=V_s=25$, the Chern numbers of the lowest three bands are calculated as $C_1=1$, $C_2=-1$ and $C_3=1$.

Next, we outline the method for obtaining the projection of the soliton state onto the linear Bloch bands. By discretizing the $x$ space and writing ${\partial^2\psi(x,t)}/{\partial x^2}=(\psi_j(x+dx,t)-2\psi(x,t)+\psi(x-dx,t))/(dx^2)$, we rewrite $H_0$ in Eq.~(\ref{H0xt}) in a discretized form. Then, we diagonalize this discretized instantaneous Hamiltonian to obtain the adiabatic eigenstate $|n\rangle$ of the $n_{th}$ level at every instant of time $t$. By projecting the renormalized wave-function $\frac{\Psi_j}{\sqrt{\langle\Psi_j|\Psi_j\rangle}}$ ($j=1,2$) onto the $n_{th}$ level, we obtain the level occupation $p_{j,n}(t)=\frac{|\langle n|\Psi_j\rangle|^2}{\langle\Psi_j|\Psi_j\rangle}$ at time $t$. For gapped cases, the instantaneous occupation of each soliton component on the $\alpha$th band is given by
\begin{align}
	P_\alpha=\sum_{n=(\alpha-1)N_L+1}^{\alpha N_L} p_n,
\end{align}
with $N_L$ is the system length. In Fig.~{\ref{Fig:FigApp1}}, we project $\Psi_j(t)$ ($j=1,2$) onto the Bloch bands, and plot the occupations on the lowest three bands, $P_{\alpha}$ ($\alpha=1,2,3$), as a function of time.

\section{Lagrangian Variational Approach}\label{AppendixC}

In this section, we present details for (i) the derivations of the Lagrangian $\mathbb{L}$ associated with GPE~(\ref{Eq:GPE}) of the main text, and (ii) the derivations of Eq.~(\ref{Eq:EqOfMotion}) for the four variational parameters.

The Lagrangian variational method has been widely applied in one-dimensional single-component~\cite{Victor1996,Franco1999,Andrea2001,Khawaja2002,Qidong2022,Cao2024} and two-component BECs~\cite{Navarro2009,Xu2019}. For the GPE~(\ref{Eq:GPE}) in the main text, the Lagrangian is derived as  
\begin{equation}
\mathbb{L}=\int_{-\infty}^{+\infty}\mathcal{L}dx,\label{L}
\end{equation}
with the Lagrangian density 
\begin{eqnarray}
	\mathcal{L}&=&\sum_{j=1,2}\left[\frac{i}{2}\left(\phi_j^\star\frac{\partial}{\partial t}\phi_j-\phi_j\frac{\partial}{\partial t}\phi_j^\star\right)-\frac{1}{2}\left|\frac{\partial\phi_j}{\partial x}\right|^2+\frac{1}{2}g|\phi_j|^4\right]\nonumber\\
	&&-\mathcal{V}_p(x+\frac{d_r}{2},t)|\phi_1|^2-\mathcal{V}_p(x-\frac{d_r}{2},t)|\phi_2|^2\nonumber\\
	&&+g_{12}|\phi_1|^2|\phi _2|^2.\label{Eq:LdensityAppen}
\end{eqnarray}
We have double-checked that the Lagrangian in Eq.~(\ref{Eq:LdensityAppen}) is valid by plugging Eq.~(\ref{Eq:LdensityAppen}) into Euler-Lagrange equation and then re-covering GP Eq.~(\ref{Eq:GPE}) in the main text. After inserting the variational ansatz in Eq.~(\ref{Eq:ansatz}) into Eqs.~(\ref{L}) and (\ref{Eq:LdensityAppen}), we obtain 
\begin{equation}
	\mathbb{L}=L_1+L_2+L_3+L_4+L_5, \label{Eq:LdensityAppenEach}
\end{equation}
where the term $L_1$ is referred as to the dynamical term, reading
\begin{eqnarray}
	L_1&=&\sum_{j=1,2}\frac{i}{2}\int_{-\infty}^{+\infty}\left(\phi_j^\star\frac{\partial \phi_j}{\partial t}-\phi_j\frac{\partial \phi_j^\star}{\partial t}\right)dx\nonumber\\
	&=&A^2\left[\left(\frac{\partial x_c}{\partial t}\right)^2+\left(\frac{\partial x_r}{\partial t}\right)^2\right]\nonumber\\
	&+&\frac{A^2}{8}\sum_{j=1,2}\left[\left(\frac{\partial \sigma_j}{\partial t}\right)^2-\sigma_j\frac{\partial^2 \sigma_j}{\partial t^2}\right].
\end{eqnarray}
The term $L_2$ in Eq.~(\ref{Eq:LdensityAppenEach}), which is associated with the kinetic energy of the model system, can be calculated as
\begin{eqnarray}
	L_2&=&-\sum_{j=1,2}\frac{1}{2}\int_{-\infty}^{+\infty}\left|\frac{\partial \phi_j}{\partial x}\right|dx\nonumber\\
	&=&-\frac{A^2}{2}\left[\left(\frac{\partial x_c}{\partial t}\right)^2+\left(\frac{\partial x_r}{\partial t}\right)^2\right]\nonumber\\
	&-&\frac{A^2}{8}\sum_{j=1,2}\left[\left(\frac{\partial \sigma_j}{\partial t}\right)^2+\frac{1}{\sigma_j^2}\right].
	\end{eqnarray}
The nonlinear term $L_3$ in Eq. (\ref{Eq:LdensityAppenEach}) represents the intra-atomic interaction energy, which is evaluated as
\begin{eqnarray}
	L_3&=&\frac{g}{2}\int_{-\infty}^{+\infty}\left|\phi_1\right|^4dx+\frac{g}{2}\int_{-\infty}^{+\infty}\left|\phi_2\right|^4dx\nonumber\\
	&=&\frac{gA^4}{8\sqrt{2\pi}\sigma_1}+\frac{gA^4}{8\sqrt{2\pi}\sigma_2}.
\end{eqnarray}
The nonlinear term $L_4$ in Eq. (\ref{Eq:LdensityAppenEach}) describes the intra-atomic interaction, i.e., 
\begin{eqnarray}
	L_4&=&g_{12}\int_{-\infty}^{+\infty}\left|\phi_1\right|^2\left|\phi_2\right|^2dx\nonumber\\
	&=&\frac{g_{12}A^4}{4\sqrt{\pi}\sqrt{\sigma_1^2+\sigma_2^2}}e^{-\frac{4x_r^2}{\sigma_1^2+\sigma_2^2}}.
\end{eqnarray}
The last term $L_5$ is associated with the spin-dependent Thouless pumps, given by
\begin{eqnarray}
	L_5&=&\frac{V_1}{2}\int_{-\infty}^{+\infty}\cos\left[4\pi\left(x+\frac{d_r}{2}\right)\right]|\phi_1|^2dx\nonumber\\
	&+&\frac{V_2}{2}\int_{-\infty}^{+\infty}\cos\left[2\pi\left(x+\frac{d_r}{2}\right)-2\nu t\right]|\phi_1|^2dx\nonumber\\
	&+&\frac{V_1}{2}\int_{-\infty}^{+\infty}\cos\left[4\pi\left(x-\frac{d_r}{2}\right)\right]|\phi_2|^2dx\nonumber\\
	&+&\frac{V_2}{2}\int_{-\infty}^{+\infty}\cos\left[2\pi\left(x-\frac{d_r}{2}\right)-2\nu t\right]|\phi_2|^2dx\nonumber\\
	&=&\frac{V_1A^2}{4}\cos\left[4\pi(x_c+x_r+\frac{d_r}{2})\right]e^{-4\pi^2\sigma_1^2}\nonumber\\
	&+&\frac{V_2A^2}{4}\cos\left[2\pi(x_c+x_r+\frac{d_r}{2})-2\nu t\right]e^{-\pi^2\sigma_1^2}\nonumber\\
	&+&\frac{V_1A^2}{4}\cos\left[4\pi(x_c-x_r-\frac{d_r}{2})\right]e^{-4\pi^2\sigma_2^2}\nonumber\\
	&+&\frac{V_2A^2}{4}\cos\left[2\pi(x_c-x_r-\frac{d_r}{2})-2\nu t\right]e^{-\pi^2\sigma_2^2}.
\end{eqnarray}

Summing up, the complete expression of the Lagrangian in (\ref{Eq:LdensityAppenEach}) can be obtained as
\begin{eqnarray}
	\mathbb{L}&=&\frac{A^2}{2}\left[\left(\frac{\partial x_c}{\partial t}\right)^2+\left(\frac{\partial x_r}{\partial t}\right)^2\right]-\frac{A^2}{8}\sum_{j=1,2}\sigma_j\frac{\partial^2 \sigma_j}{\partial t^2}\nonumber\\
	&+&\frac{g_{12}A^4}{4\sqrt{\pi}\sqrt{\sigma_1^2+\sigma_2^2}}e^{-\frac{4x_r^2}{\sigma_1^2+\sigma_2^2}}+\sum_{j=1,2}\left(\frac{gA^4}{8\sqrt{2\pi}\sigma_1}-\frac{A^2}{8\sigma_1^2}\right)\nonumber\\
	&+&\frac{A^2}{4}\left\{V_1 e^{-4\pi^2\sigma_1^2}\cos\left[4\pi\left(x_c+x_r+\frac{d_r}{2}\right)\right]\right.\nonumber\\
	&+&\left.V_2e^{-\pi^2\sigma_1^2}\cos\left[2\pi\left(x_c+x_r+\frac{d_r}{2}\right)-2\nu t\right]\right.\nonumber\\
	&+&\left.V_1e^{-4\pi^2\sigma_2^2}\cos\left[4\pi\left(x_c-x_r-\frac{d_r}{2}\right)\right]\right.\nonumber\\
	&+&\left.V_2e^{-\pi^2\sigma_2^2}\cos\left[2\pi\left(x_c-x_r-\frac{d_r}{2}\right)-2\nu t\right]\right\}.\label{AppLF}
\end{eqnarray}
\begin{widetext}

Finally, the equations of motion of the variational parameters of $y_i=x_c,x_r,\sigma_1,\sigma_2$ can be obtained from the Euler-Lagrange equation $\frac{\partial \mathbb{L}}{\partial y_i}-\frac{d}{dt}\left(\frac{\partial \mathbb{L}}{\partial \dot{y}_i}\right)=0$. After straightforward algebra and some rearrangements, we obtain
	\begin{eqnarray}
		\frac{d^2}{dt^2}x_c&=&-\pi V_1\sin\left[4\pi\left(x_c+x_r+\frac{d_r}{2}\right)\right]e^{-4\pi^2\sigma_1^2}-\frac{\pi V_2}{2}\sin\left[2\pi\left(x_c+x_r+\frac{d_r}{2}\right)-2\nu t\right]e^{-\pi^2\sigma_1^2}\nonumber\\
		&-&\pi V_1\sin\left[4\pi\left(x_c-x_r-\frac{d_r}{2}\right)\right]e^{-4\pi^2\sigma_2^2}-\frac{\pi V_2}{2}\sin\left[2\pi\left(x_c-x_r-\frac{d_r}{2}\right)-2\nu t\right]e^{-\pi^2\sigma_2^2},\label{AppXc}\\
                \frac{d^2}{dt^2}x_r&=&-\frac{2g_{12}A^2x_r}{\sqrt{\pi}(\sigma1^2+\sigma_2^2)^{3/2}}e^{-\frac{4x_r^2}{\sigma_1^2+\sigma_2^2}}\nonumber\\
                &-&\pi V_1\sin\left[4\pi(x_c+x_r+\frac{d_r}{2})\right]e^{-4\pi^2\sigma_1^2}-\frac{\pi V_2}{2}\sin\left[2\pi(x_c+x_r+\frac{d_r}{2})-2\nu t\right]e^{-\pi^2\sigma_1^2}\nonumber\\
		&+&\pi V_1\sin\left[4\pi(x_c-x_r-\frac{d_r}{2})\right]e^{-4\pi^2\sigma_2^2}+\frac{\pi V_2}{2}\sin\left[2\pi(x_c-x_r-\frac{d_r}{2})-2\nu t\right]e^{-\pi^2\sigma_2^2},\label{AppXr}\\
                \frac{d^2}{dt^2}\sigma_1&=&\frac{2}{\sigma_1^3}-\frac{gA^2}{\sqrt{2\pi}\sigma_1^2}+\frac{16g_{12}A^2x_r^2\sigma_1}{\sqrt{\pi}(\sigma_1^2+\sigma_2^2)^{5/2}}e^{-\frac{4x_r^2}{\sigma_1^2+\sigma_2^2}}-\frac{2g_{12}A^2\sigma_1}{\sqrt{\pi}(\sigma_1^2+\sigma_2^2)^{3/2}}e^{-\frac{4x_r^2}{\sigma_1^2+    \sigma_2^2}}
                \nonumber\\
                &-&16\pi^2V_1\sigma_1\cos\left[4\pi(x_c+x_r+\frac{d_r}{2})\right]e^{-4\pi^2\sigma_1^2}\nonumber\\
                &-&4\pi^2V_2\sigma_1\cos\left[2\pi(x_c+x_r+\frac{d_r}{2})-2\nu t\right]e^{-\pi^2\sigma_1^2},\label{Appsigma1}\\
	\frac{d^2}{dt^2}\sigma_2&=&\frac{2}{\sigma_2^3}-\frac{gA^2}{\sqrt{2\pi}\sigma_2^2}+\frac{16g_{12}A^2x_r^2\sigma_2}{\sqrt{\pi}(\sigma_1^2+\sigma_2^2)^{5/2}}e^{-\frac{4x_r^2}{\sigma_1^2+\sigma_2^2}}-\frac{2g_{12}A^2\sigma_2}{\sqrt{\pi}(\sigma_1^2+\sigma_2^2)^{3/2}}e^{-\frac{4x_r^2}{\sigma_1^2+\sigma_2^2}}
	\nonumber\\
		&-&16\pi^2V_1\sigma_2\cos\left[4\pi(x_c-x_r-\frac{d_r}{2})\right]e^{-4\pi^2\sigma_1^2}\nonumber\\
		&-&4\pi^2V_2\sigma_2\cos\left[2\pi(x_c-x_r-\frac{d_r}{2})-2\nu t\right]e^{-\pi^2\sigma_1^2},\label{Appsigma2}
	\end{eqnarray}
	\end{widetext}
Equations (\ref{AppXc})-(\ref{Appsigma2}) form a closed set of equations, the solutions of which give the time evolution of the vector soliton.

We can interpret the dynamics of $x_c$, $x_r$, $\sigma_1$ and $\sigma_2$ of the vector solitons in Eqs.~(\ref{AppXc})-(\ref{Appsigma2}) in terms of the motion of a classical particle of mass $m_{\text{eff}}$ subjected to an external potential of $V_{\text{ex}}$ as $m^i_{\text{eff}}d^2y_i/dt^2=-\partial V_{\text{ex}}/\partial y_i$ with $y_1=x_c$, $y_2=x_r$, $y_3=\sigma_1$ and $y_4=\sigma_2$. In this end, we rewrite the Lagrangian in Eq.~(\ref{AppLF})  in the form of $\mathbb{L}=T-V_{\text{ex}}$ with $T$ and $V_{\text{ex}}$ denoting the generalized kinetic energy and the generalized potential energy, respectively. In more details, the concrete expressions of $T$ and $V_{\text{ex}}$ can are respectively written as follows
\begin{equation}
 \!\!\!    T=\frac{A^2}{2}\left[\left(\frac{dx_c}{dt}\right)^2+\left(\frac{dx_r}{dt}\right)^2\right]-\sum_{j=1,2}\frac{\sigma_jA^2}{8}\frac{d^2\sigma_j}{dt^2},
\end{equation}
and
\begin{eqnarray}
	V_\textrm{ex}&=&\sum_{j=1,2}\left(\frac{A^2}{8\sigma_j^2}-\frac{gA^4}{8\sqrt{2\pi}\sigma_j}\right)-\frac{g_{12}A^4e^{-4x_r^2/(\sigma_1^2+\sigma_2^2)}}{4\sqrt{\pi(\sigma_1^2+\sigma_2^2)}} \nonumber\\
	\nonumber\\
	&-&\frac{A^2}{4}\left\{V_1 e^{-4\pi^2\sigma_1^2}\cos[4\pi(x_c+x_r+\frac{d_r}{2})]\right.\nonumber\\
	&+&\left.V_2e^{-\pi^2\sigma_1^2}\cos[2\pi(x_c+x_r+\frac{d_r}{2})-2\nu t]\right.\nonumber\\
	&+&\left.V_1e^{-4\pi^2\sigma_2^2}\cos[4\pi(x_c-x_r-\frac{d_r}{2})]\right.\nonumber\\
	\!\!\!\!\!\!\!\!&+&\left.V_2e^{-\pi^2\sigma_2^2}\cos[2\pi(x_c-x_r-\frac{d_r}{2})-2\nu t]\right\}.\label{Veff}
\end{eqnarray}
As a result, the equations of motion for $x_c$ in Eq.~(\ref{AppXc}), $x_r$ in Eq.~(\ref{AppXr}), $\sigma_1$ in Eq.~(\ref{Appsigma1}) and $\sigma_2$ in Eq.~(\ref{Appsigma2}) can be obtained by using the effective potential in Eq.~(\ref{Veff}). Thus we obtain \begin{eqnarray}
m^i_{\text{eff}} \frac{d^2{y_i}}{dt^2}&=&-\frac{\partial V_\textrm{ex}}{\partial y_i},\label{AppEqOfMotion}
\end{eqnarray}
with the effective mass being $m_{\text{eff}}=[A^2,A^2,A^2/8,A^2/8]$. 

Equations~(\ref{AppXc})-(\ref{Appsigma2}) allow for a simple picture for the dynamics of $x_c$, $x_r$, $\sigma_1$ and $\sigma_2$ when we associate
with them the classical motion of a fictitious particle with the effective mass $m_{\text{eff}}=[A^2,A^2,A^2/8,A^2/8]$ in Eq.~(\ref{AppEqOfMotion}).

\bibliography{Reference}
	
\end{document}